\documentclass[prd,nofootinbib,onecolumn,superscriptaddress]{revtex4}

\usepackage{amsmath,amstext,amsbsy,amsopn,amsthm,amscd,amsfonts,amssymb}
\usepackage{bm}
\usepackage{epsfig}
\usepackage{graphpap}
\usepackage{mathrsfs}
\usepackage{setspace}

\newcommand{\para}{\vspace{+1.5ex}}

\theoremstyle{definition}

\theoremstyle{remark}

\newcommand{\REMOVE}[1]{}

\newcommand{\begeq}{\begin{equation}}
\newcommand{\eneq}{\end{equation}}
\newcommand{\beq}{\begin{equation}}
\newcommand{\eeq}{\end{equation}}
\newcommand{\beqa}{\begin{eqnarray}}
\newcommand{\eeqa}{\end{eqnarray}}

\newcommand{\equa}[1]{\begin{equation}#1\end{equation}}
\newcommand{\equasn}[1]{\begin{align}#1 \end{align}}
\newcommand{\equas}[1]{\begin{equation}\begin{array}{l}#1\end{array}\end{equation}}
\newcommand{\matr}[1]{\left( \begin{matrix} #1 \end{matrix} \right)}

\newcommand{\nonum}{\nonumber \\}

\newcommand{\epsq}{\epsilon^2}
\newcommand{\eps}{\epsilon}
\newcommand{\del}{\delta}
\newcommand{\deln}[1]{\delta^{#1}}

\newcommand{\GeV}{\text{GeV}}

\newcommand{\bifundrep}{{\bf{( 5,\bar 5)}}}
\newcommand{\antibifundrep}{{\bf{(\bar 5, 5)}}}
\newcommand{\firstfive}{{\bf{(5,1)}}}
\newcommand{\firstfiveb}{{\bf{(\bar 5,1)}}}
\newcommand{\firstten}{{\bf{(10,1)}}}
\newcommand{\secfive}{{\bf{(1,5)}}}
\newcommand{\secfiveb}{{\bf{(1,\bar 5)}}}
\newcommand{\secten}{{\bf{(1,10)}}}
\newcommand{\h}{{\bf{(5,1)}}}

\newcommand{\phibard}{{\bar{\Phi}}_d}
\newcommand{\phibart}{{\bar{\Phi}}_t}
\newcommand{\phibar}{{\bar{\Phi}}}
\newcommand{\phit}{\Phi_t}
\newcommand{\phid}{\Phi_d}
\newcommand{\fbarpr}{{\bf \bar F^\prime}}
\newcommand{\fbar}{{\bf \bar{F}}}
\newcommand{\T}{{\bf T}}
\newcommand{\tpr}{{\bf T^\prime}}
\newcommand{\hbarpr}{\bar h^\prime}

\newcommand{\hpr}{h^\prime}
\newcommand{\barh}{\bar h}
\newcommand{\zn}[1]{Z_n(#1)}
\newcommand{\znpr}{Z_n^\prime}

\newcommand{\vev}[1]{\langle #1 \rangle}

\newcommand{\mpl}{M_\text{Pl}}
\newcommand{\mgut}{M_\text{GUT}}
\newcommand{\suf}{SU(5) }
\newcommand{\sufns}{SU(5)}
\newcommand{\su}[1]{SU(#1) }

\newcommand{\suff}{SU(5)$\times$\;SU(5) }

\newcommand{\gut}{GUT }

\newcommand{\smggns}{ SU(3)$\times$\;SU(2)$\times$\;U(1)}
\newcommand{\order}{\mathcal{O}}

\newcommand{\diag}[1]{\text{diag}(#1)}

\begin{document}
\title%
{The flavor of product-group GUTs}
 \author{Yehonathan Segev}\email{yehonathan.segev@weizmann.ac.il}
 \affiliation{Physics Department,
Technion--Israel Institute of Technology, Haifa
32000, Israel}\affiliation{Department of
Condensed Matter Physics, Weizmann Institute of
Science, Rehovot 76100, Israel}
\author{Yael Shadmi}\email{yshadmi@physics.technion.ac.il}
\affiliation{Physics Department, Technion--Israel Institute of
  Technology, Haifa 32000, Israel}

\begin{abstract}
The doublet-triplet splitting problem can be
simply solved in product-group GUT models,
using a global symmetry that distinguishes the
doublets from the triplets. Apart from giving
the required mass hierarchy, this ``triplet
symmetry'' can also forbid some of the triplet
couplings to matter. We point out that, since
this symmetry is typically
generation-dependent, it gives rise to
non-trivial flavor structure. Furthermore,
because flavor symmetries cannot be exact, the
triplet-matter couplings are not forbidden then
but only suppressed. We construct models in
which the triplet symmetry gives acceptable
proton decay rate and fermion masses. In some
of the models, the prediction $m_b\sim m_\tau$
is retained, while the similar relation for the
first generation is corrected. Finally, all
this can be accomplished with triplets somewhat
below the GUT scale, supplying the right
correction for the standard model gauge
couplings to unify precisely.
\end{abstract}

\maketitle
\thispagestyle{empty}
\section{Introduction}
The doublet-triplet splitting problem can be elegantly solved in
Grand Unified Theories (GUTs) based on semi-simple GUT
groups~\cite{{Barbieri94},{Barr96},{Witten01},{Dine02}}. If the
standard-model (SM) Higgses originate from GUT fields that
transform under different factors of the GUT group, these theories
can accommodate a global symmetry, which we will refer to in the
following as a ``triplet symmetry'', that allows a triplet mass at
the GUT scale while forbidding a doublet mass. Furthermore, the
triplet symmetry may also forbid some of the triplet couplings to
standard-model (SM) matter fields~\cite{Witten01,Dine02}.
This eliminates dangerous
contributions to the proton decay rate, so that GUT-scale triplets
are consistent with current bounds on proton decay, unlike in
minimal SU(5). In fact, even triplets below the
GUT scale are allowed, and one can construct models in which the
triplets are around 10$^{15}$ GeV, so that they provide precisely
the right threshold correction for successful coupling
unification~\cite{Rakshit03,{Murayama01}}.

As mentioned above, the standard-model Higgses
transform under different group factors in these models.
Likewise, the SM matter fields can transform
under different GUT group factors. This has two
immediate consequences for the fermion mass
matrices. First, some fermion mass terms
involve fields transforming under different
group factors. Because of the GUT gauge
symmetry, such a mass term must come from a
higher-dimension term that includes GUT
breaking fields, and is therefore suppressed by
a power of $\eps=\mgut/\mpl$. Thus, the GUT
gauge symmetry automatically generates some
non-trivial fermion mass textures
\cite{Dine02}.
Second, in such models the triplet symmetry is {\it necessarily}
horizontal---or generation-dependent \cite{Rakshit03}.
It therefore dictates a non-trivial structure
of fermion mass matrices on top of the texture
generated by the GUT gauge symmetry.

In this paper, we will study the flavor structure of \suff models.
As we will see, the triplet symmetry can generate viable mass
matrices. The picture that emerges is very attractive. The same
global symmetry generates a doublet-triplet mass hierarchy,
suppresses the triplet contribution to proton decay, and gives
viable fermion masses. In addition, some of the models
{\emph{partially}} break the usual GUT `` Yukawa unification'', so
that the successful relation  $m_b\sim m_\tau$ is maintained while
a similar relation for the first generation is avoided.

Since the triplet symmetry is generation
dependent, it must be broken---otherwise some
fermion mass splittings and/or mixings vanish
\cite{Nir92,{Nir93}}. Therefore, triplet-matter
couplings that would have been forbidden had
the symmetry been exact, are no longer zero. In
fact, the proton decay rate may even be larger
in some \suff models compared to minimal \sufns, since
the triplet couplings to matter may be
\emph{enhanced} relative to the doublet
couplings. Therefore, apart from checking the
flavor parameters of the models, we must check
the proton decay rate.

The paper is organized as follows. We start
(section~\ref{basics}) with a brief review of
the basics of \suff models. In
section~\ref{effect_of_triplet_sym} we list the
possible Higgs sectors of the models,
and discuss the effects of the triplet symmetry on flavor
and on proton decay.
We find that \textbf{(a)} the
up sector mass matrix hierarchies, \textbf{(b)}
the mass ratio of a lepton and corresponding
down-sector quark, \textbf{(c)} the ratio of
the triplet Yukawa couplings and corresponding
doublet coupling, are all governed by a single
parameter.
Model building is then reduced to finding a
parameter  that satisfies  flavor
and proton decay constraints. In
section~\ref{results} we apply this to
specific models.
Finally,
motivated by gauge coupling unification, we
complete our analysis in section~\ref{lightTriplets}
by looking into the
possibility of having the triplets
at $10^{14}$--$10^{15}$GeV. We
summarize our results in section~\ref{summary}.
In the Appendix we discuss a model that is
ruled out only by the combined constraints from
proton decay and flavor.
%
\section{Basic Structure}
\label{basics}
We consider supersymmetric models with gauge
group $G=$\sufns$_1\times$\sufns$_2$, that have
an additional $Z_n$ global symmetry. Refs.
\cite{Barbieri94,Barr96,Witten01,Dine02} show
how this setup may be used to account for the
doublet-triplet splitting. The basic scenario
of \cite{Dine02} is briefly described below. We consider $G$
breaking down to \smggns$\times\znpr$ by the
following VEVs of bifundamental fields $\phit$
and $\phid$ [$\phibart$ and $\phibard$]
transforming as $\bifundrep$
[$\antibifundrep$],
\equa{
\begin{array}{lclcl}
\label{bifundVEVs}
\vev{\phit}&=&\vev{\phibart}&=&\diag{v_t,v_t,v_t,0,0}\ ,\\
\vev{\phid}&=&\vev{\phibard}&=&\diag{0,0,0,v_d,v_d}\ .
\end{array}
}
As shown in~\cite{Dine02}, these VEVs may
correspond to exact or approximate flat
directions with $v_t\sim v_d\sim\mgut$, so we
will take $v_t=v_d=v$ for simplicity.
Since the SM gauge group
is contained in the diagonal subgroup of $G$,
 the low energy couplings
$\alpha_1,\alpha_2,\alpha_3$ unify even when the
couplings of the two \suf factors are
different, and charge quantization is
maintained.

The MSSM Higgs fields, now embedded in a \gut
multiplet, are taken to transform under
different group factors. For example, let us
consider $h$ and $\hbarpr$, transforming as
$\firstfive$ and $\secfiveb$. The
superpotential terms
\equa{
h\phibart\hbarpr\ , \ \ \ \
h\phibard\hbarpr\ ,
}
give a triplet mass
term and a doublet mass term respectively.
Clearly, if
\equa{
\zn{\phit}-\zn{\phid}=\zn{\phibard}-\zn{\phibart}\neq
0\ ,
}
then the doublet and triplet masses cannot be
allowed simultaneously.

The bifundamental VEVs~(\ref{bifundVEVs}) leave a combination of
$Z_n$ and the hypercharge of \sufns$_1$
unbroken. This combination,
\equa{
\label{znpr_def} \znpr=Y_1^k\times Z_n\ ,
}
with some integer $k$ and with
\equa{
Y_1=\diag{e^{-2i\frac{2\pi}{n}},e^{-2i\frac{2\pi}{n}},e^{-2i\frac{2\pi}{n}},
e^{+3i\frac{2\pi}{n}},e^{+3i\frac{2\pi}{n}}}\
,
}
is the triplet symmetry. The Higgs triplets and
doublets transform differently under $\znpr$,
\equas{
    \label{higgsZnTrans}
    \matr{ h_t, & h_d }
    \rightarrow
    \matr{ e^{i\frac{2\pi}{n}(\zn{h}-2k)} h_t, &
    e^{i\frac{2\pi}{n}(\zn{h}+3k)}
    h_d}\ , \;
    \\
    \matr{ \hbarpr_t, & \hbarpr_d }
    \rightarrow
    \matr{ e^{i\frac{2\pi}{n}{\zn{\hbarpr}}}\hbarpr_t,
    & e^{i\frac{2\pi}{n}{\zn{\hbarpr}}} \hbarpr_d
    }\ .
}
The matter fields, too, may be split between
the two group factors.
This has immediate consequences for the quark
and lepton masses~\cite{{Witten01}, Dine02}.
Denoting,
\equas{
    \T\sim\bf{(10,1)}\ ,\ \  \tpr\sim\bf{(1,10)}\ ,\\
    \fbar\sim\bf{(5,1)}\ ,\ \  \fbarpr\sim\bf{(1,\bar
    5)}\ ,
}
with the Higgs field $h\sim\h$,
the following mass terms may arise (depending
on the matter content of the specific model)
\equasn{
\T\T h &\sim \bf{(10,1)(10,1)(5,1)}\ , \nonumber \\
    \frac{1}{\mpl^2}\tpr\T h \Phi \Phi&\sim
    \frac{1}{\mpl^2} \bf{(1,10)(10,1)(5,1)(5,\bar
    5)(5,\bar5)}\ ,
}
where $\mpl$ is the Planck scale.
Some mass terms are then suppressed by powers of
\equa{
\eps=\frac{v}{\mpl}\sim\frac{\mgut}{\mpl}\sim
10^{-2}\ ,
}
and a fermion mass hierarchy is generated.
We refer to this structure of the
mass matrices as $\eps$-dependence.

 Models that have matter generations transforming under
different group factor have some Yukawa
couplings mediated by $\phit$ ($\phibart$) and
some mediated by $\phid$ ($\phibard$). Since
the two bifundamentals carry different $Z_n$
charges, the discrete symmetry distinguishes
Yukawa couplings that would have been equal in
minimal \sufns.
This may partially break the
lepton-down sector Yukawa unification \cite{Dine02}, and
suppress the couplings of the Higgs triplet to matter,
thus suppressing proton decay mediated by
Higgsino exchange \cite{Witten01,Dine02}.

A further consequence of splitting matter between
the two group factors is that in such models
$\znpr$ is necessarily horizontal.
Suppose a model
has {\bf 10}s coming from $\T\sim{\bf (10,1)}$
and $\tpr\sim{\bf (1,10)}$.
For $\znpr=Z_n\times Y^k_1$ to be generation blind, we
have to require
\equa{
\znpr(Q)=\znpr(Q^\prime)\ ,
}
 where $Q$ and
$Q^\prime$ are the \su{2} doublet superfields
coming from $\T$ and $\tpr$ respectively. From
the definition of $\znpr$ it follows that
fields coming from $\tpr$ have the same $\znpr$
charge, so that
\equa{
\znpr({u^c}^\prime)=\znpr(Q^\prime)\ ,
}
with ${u^c}^\prime$ the up sector \su{2}
singlet contained in $\tpr$. However, fields
coming from $\T\sim{\bf (10,1)}$ are then
distinguished by $\znpr$,
\equa{
 \znpr({u}^c)=\znpr(Q)-kY_1(Q)+kY_1(u^c)\neq\znpr({u^\prime}^c)\ ,
}
so the triplet symmetry is generation dependent.
Therefore, it must be broken, otherwise degenerate quarks or
zero mixing angles result~\cite{Nir92,Nir93}.
Thus, the mass matrices acquire an additional
hierarchy, and Yukawa couplings that would have
been forbidden if the symmetry were exact are
now allowed.

We assume then that $\znpr$ is broken by the
VEV of a gauge singlet field $S$, with charge
$\zn{S}=s$, and that the standard-model Yukawa
couplings originate from higher-dimension terms
involving some power of $S$, which is dictated
by the triplet symmetry. Below the scale
$\langle S\rangle$, the resulting Yukawa
couplings acquire a flavor hierarchy
parameterized by powers of
$\delta\equiv{\vev{S}}/{\mpl}$, on top of their
$\eps$-dependence. Since the bifundamentals
carry zero $\znpr$ charge, the  $\eps$'s are
$\znpr$ neutral, and the $\eps$- and
$\del$-dependences can be studied separately.
We also assume that $\delta$ is not much
smaller than $\sin\theta_c$, and will take
$\delta\sim0.1$ for concreteness. We stress
that in our analysis we ignore order one
coefficients and stick to counting powers of
$\eps$'s and $\del$'s.

For later convenience we define the parameter
$w\in \{0,1,\dots, n-1\}$, such that
\equa{
ws=\zn{\phit}-\zn{\phid}\ .
}
The doublet triplet splitting condition becomes
$ws\neq 0$. Consider the ratio of two Yukawa
couplings that would have been equal in minimal
\sufns. If one involves $\phit$, and the
other $\phid$, their ratio is  $\del^w$.
 This
parameter will enter in the suppression of the
triplet couplings relative to the doublet
couplings, as well as in the ratio of lepton-
and down-quark masses, as we show in section
\ref{effect_of_triplet_sym}.

\section{The Effect of the Triplet Symmetry}
\label{effect_of_triplet_sym}

\subsection{Higgses and model classification}
\label{extList}
As discussed above, the MSSM matter and Higgs
fields can transform under either one of the
\sufns's. Following Ref.~\cite{Dine02}, we list
below all possible assignments such that {\bf
(a)} the model is anomaly free and {\bf (b)}
the top and up type Higgs transform under the
same \sufns, so that the top Yukawa is
renormalizable. In some cases we add additional
$\firstfive+\secfiveb$ to cancel anomalies.
\newcounter{modelClass}
\newcounter{modelSubClass}
\newcounter{modelssclass}
 \numberwithin{modelSubClass}{modelClass}
\begin{list}{\Alph{modelClass}.}{\usecounter{modelClass}}
\item In this class the MSSM Higgses come from $h\sim \firstfive$ and
$\hbarpr\sim\secfiveb$,  with no additional
$\firstfive$ and $\secfiveb$ pairs. The triplet
mass term $h_t\phibart\hbarpr_t$ is assumed to
be $\znpr$ neutral so the triplets acquire
$\mgut$ mass. (We consider lower triplet
mass consistent with gauge coupling unification
in section~\ref{lightTriplets}.) Consequently
the doublet mass term $h_d\phibard\hbarpr_d$
has a $Z_n$ charge of $+ws$, so the doublet
mass is suppressed by $\deln{n-w}$. We can then
arrange for the doublets to be at the
electroweak scale by taking $n$ sufficiently
large. The MSSM matter fields come from
\begin{list}{\Alph{modelClass}.\arabic{modelSubClass}:}
{\usecounter{modelSubClass}}
\item $3\times \firstfiveb + 2\times \firstten
+\secten$\ ,
\item $ 2\times [\firstfiveb+\secten] + \firstten+\secfiveb$\ .
\end{list}
\item The model contains
$h\firstfive,\barh \firstfiveb,\hpr\secfive$
and $\hbarpr\secfiveb$. Of these, the MSSM
Higgses come from $h$ and $\hbarpr$; $\hpr$ and
$\barh$ do not couple to the MSSM fields (this
can be ensured by imposing an additional
symmetry). $h$ and $\hbarpr$ gain mass through
a mutual mass term, and the Higgs-sector
spectrum is as in~A.
The MSSM matter generations come from
\begin{list}{\Alph{modelClass}.\arabic{modelSubClass}:}{\usecounter{modelSubClass}}
\item $3\times [\firstfiveb+\firstten]$\ ,
\item
$2\times[\firstfiveb+\firstten]+\secfiveb+\secten$\
,
\item
$\firstfiveb+\firstten+2\times[\secfiveb+\secten]$\ .
\end{list}
\item This model contains  $h,\bar h,h^\prime$
and $\hbarpr$, with all triplets heavy.
The matter fields are as in models B.
The Higgs mass terms are
(neglecting $\order(1)$ coefficients):
\equa{
\label{fourHiggsMassTerms}
\begin{array}{ccl}
 \Delta W &=& {S^A\over\mpl^{A-1}} \,  h\barh
+ {S^{n-A}\over\mpl^{n-A-1}}\, \hpr\hbarpr
+\vev{\phibart}h \hbarpr +\vev{\phit}\hpr \barh
+{S^w\over\mpl^{w}} \, \vev{\phibard}\, h
\hbarpr
 + {S^{n-w}\over\mpl^{n-w}}\,
\vev{\phid}\hpr \barh \nonumber\\
 &\to& \delta^A \mpl
h\barh +\delta^{n-A} \mpl\hpr\hbarpr
+\vev{\phibart}h \hbarpr +\vev{\phit}\hpr \barh
+\delta^w \vev{\phibard}h \hbarpr
  +
\delta^{n-w}\vev{\phid}\hpr \barh \ ,\nonumber
\end{array}
}
where $A\in Z_n$ is a free parameter, and we have to require
$A\neq 0$, or all Higgses, triplets and doublets alike, get $\mpl$
mass.

Inspecting the resulting triplet- and doublet-mass matrices it is
easy to see that the only acceptable choices are $A=1,2$. In this
case, one triplet pair and one doublet pair are at or above the
GUT scale, with mass $\delta^A \mpl$, and the second triplet pair
is at or just below the \gut scale, with mass
$(\eps/\delta^A)\mgut$. Finally the MSSM Higgs doublets, which are
made predominantly of $h_d$ and $\barh_d$, are light, with mass
$\delta^{n-A} \mpl$.
\end{list}

\para
We are interested in the flavor structure of
models in which the triplet symmetry is
horizontal. Models B.1 and C.1 are
not of this type. Hence, the  flavor structure
of these models must come from
another mechanism. As for the proton decay rate,
in model C.1, all Yukawa couplings come from
renormalizable terms, so the triplet and
doublet Yukawa couplings are equal as in
minimal \sufns. Therefore the proton decay rate
in this model is the same as in minimal \suf
and it is ruled out
\cite{Murayama01,Bajc02,Costa03,Wiesenfeldt04}.
Unlike model C.1, in model B.1 all dangerous
dimension-5 operators are zero and the model is
viable. We will not consider this model
further.

\para
Of the remaining models we show that models
A.1, A.2, B.2, and B.3 exhibit viable flavor
parameters and suppressed proton decay rate.
Models C.2 and C.3 are shown to be more
severely constrained by proton decay. Model
C.2 is shown to be ruled out by the combination
of flavor and proton decay constraints.

\subsection{Flavor}
We now turn to a systematic analysis of the
models starting with the general flavor
structure, and continuing with down lepton
splitting in section~\ref{e-d_splitting} and
proton decay in~\ref{proton_decay}.

Consider first the relative hierarchy and
mixing between fields that come from ${\bf 10}$s
charged under the same \sufns, say
$\tpr_1,\tpr_2$ (the indices denote
generations), as in models A.2, B.3, and C.3.
It is easy to show that
\equa{
\label{reduction2gen}
\frac{m_{u_{11}}}{m_{u_{22}}}\sim\left(
\frac{m_{u_{12}}}{m_{u_{22}}}\right)^2.
}
These models therefore imply that
\equa{
\label{tpr_tpr_mixing}
\frac{m_u}{m_c}\sim \left(V_{us}\right)^2,
}
which is off by 10--40. Thus
the up-quark mass is generically too high in
these models, and must be further suppressed.
The suppression, however, cannot be done with
an additional $Z_n$ symmetry because
\eqref{reduction2gen} will still hold.

The corresponding equality in models A.1, B.2,
and C.2, that have the ${\bf 10}$s  ${\bf \T_2}$
and ${\bf \T_3}$, is more successful
\equa{
\frac{m_c}{m_t}\sim \left(V_{cb}\right)^2,
}
which is consistent within our order of magnitude analysis.

Now let us consider the flavor parameters of
two generations that have ${\bf 10}$s
transforming under different \sufns 's--- $\T$
and $\tpr$. Specifically, let's consider the up
sector mass ratio: ${m_{Q^\prime
{u^c}^\prime}}/{m_{Q u^c}}$ where, $Q^\prime,
{u^c}^\prime \in \tpr$, and $Q,u^c \in \T$, and
the mixing of the two generations ${m_{Q^\prime
x}}/{m_{Q x}}$, where $x=u^c
,{u^c}^\prime,d^c,{d^c}^\prime$. We define the
parameters, $z$ and $r$, by
\equa{
\label{z_and_r_def}
 \frac{m_{Q^\prime
{u^c}^\prime}}{m_{Q u^c}}\sim \eps\delta^z\
, \quad\frac{m_{Q^\prime x}}{m_{Q x}} \sim
\eps^\#\delta^r,
}
where $\eps^\#= \eps^0,\eps^{\pm 1},\eps^{\pm
2}$ depending on $x$ and the Higgses of the
specific model. A priori, the ambiguity in the
power of $\eps$ makes the choice of $r$
ambiguous. However, in all our models the
leading contribution to the CKM element
$V_{u^\prime d}$, will come from Yukawa terms
that obey $V_{u^\prime d}\sim {m_{Q^\prime
x}}/{m_{Q x}} \sim \eps^0\delta^r$, so that $r$
may related to the experimental data by
$V_{u^\prime d}\sim \delta^r$.

Note also that, since the bifundamental  VEVs
are $\znpr$ neutral,
the $\delta$-dependence of any Yukawa coupling is
only determined by the $\znpr$ charges of the
low energy matter fields and Higgses and does
not depend on which bifundamental mediates the
coupling. Specifically, $r$ is determined only
by the $\znpr$ charge difference of $Q$ and
$Q^\prime$.

We can rewrite eqn.~(\ref{z_and_r_def}) as,
\equasn{
\label{deltaDependenceEqs}
 Z_n^\prime(Q^\prime)
+ Z_n^\prime({u^c}^\prime)&= Z_n^\prime(Q)+ Z_n^\prime(u^c)-zs
\nonum Z_n^\prime(Q^\prime)&=Z_n^\prime(Q)-rs \pmod{n}\ ,
}
using~\eqref{znpr_def} this neatly comes down to
\equa{
\label{flavorEqSU(5)XSU(5)} (2r-z)s=5k
\pmod{n}\ .
}
The parameters $r$ and $z$ may be related to $w$ by
noting that since the non-zero expectation
values of the bifundamentals transform
trivially under $\znpr$ then
\equa{
\left\{ \begin{array}{l}
0=k(-2)+Z_n(\Phi_t)\\
0=k(+3)+Z_n(\Phi_d)\pmod{n}\ ,
\end{array}
\right. \nonumber
}
so that
\equa{
\label{bifundamentalCharges}
ws=(\zn{\phit}-\zn{\phid})=+5k \pmod{n}\ .
}
Thus, assuming for simplicity
$\gcd[s,n]=1$,
\equa{
w=(2r-z) \pmod{n}\ .
}
Since $r$ and $z$ are input parameters that are determined
by the observed masses and mixings [see eqn.~(\ref{z_and_r_def})],
$w$ is now determined.
We show next that $w$ is also related to the down
quark-lepton mass splitting, and to the
suppression of proton decay.

\subsection{Electron-down splitting}
\label{e-d_splitting}
In any model that has different numbers of {\bf
10}s and $\bf {\bar 5}$s
coming from the first
\sufns, there is a diagonal down-sector mass
term that involves a ${\bf 10}$ and a
$\bf\bar{5}$ of different \sufns 's and is mediated
by $\Phi$ ($\phibar$). Furthermore, the mass
term of the down quark in these models,
involves $\phit$ (or $\phibart$), while the
lepton mass term involves $\phid$ (or
$\phibard$). It follows that
\equa{
\label{e-d_split_general}
\zn{m_{l}^T}=\zn{m_d} \pm (\zn{\Phi_d}-\zn{\Phi_t})\ ,
}
where $\zn{m_l^T}$ [$\zn{m_d}$] is the total $Z_n$ charge of the
lepton [down quark] mass term and the minus sign is used whenever
the mass term is mediated by $\phibar$ rather than $\Phi$. This
means that (keeping in mind the modulo-${n}$ math)
\equa{
\label{e-d_ratio} m_{l}^T\sim m_d \delta^{\pm w}\
.
}
In models A, one then gets
\equa{
\frac{m_e}{m_d}\sim\del^w\ ,
}
by taking the first generation matter
representations to be $\tpr_1,\fbar_1$. Yukawa
unification is maintained for the two heavier
generations.

One may wish to use this mechanism in models that have equal
numbers of $\fbar$s and $\T$s by taking the following generation
representations\footnote{Since we require the top mass term to be
renormalizable, all our models have the MSSM $h^U$ coming from
$h\sim \firstfive$, and since we wish to maintain $m_\tau\sim
m_b$, the third generation representations have to be
$\T_3,\fbar_3$. Thus, the above option for e-d splitting is
relevant to models B.2 and C.2. However, the latter model is
not viable (see Appendix~\ref{modelC2}).
}: $\tpr_1,\T_2,\fbar_1,
\fbarpr_2$. Although this may a-priori enable splitting the masses
of the two light generations, we could not build such models.

Splitting the down and lepton masses introduces
a constraint on $w$. If we want to arrange for
$m_e/m_d \sim 0.1$ with $\del=\order(0.1)$, we
have to take $w=1$. Other charge assignments
will result in an electron mass that is too
small. This limits the flavor parameters these
models may give. For example, in model A.1
(with the MSSM generations coming from $\tpr_1,
\T_2, \T_3, \fbar_1, \fbar_2, \fbar_3$) taking
$V_{us}\sim\del$ forces
${m_u}/{m_c}\sim\eps\del$, in  good agreement
with observation. However, in model A.2 (with
the MSSM generations coming from $\tpr_1,
\tpr_2, \T_3, \fbar_1, \fbarpr_2 \fbar_3$),
taking ${m_c}/{m_t}\sim\eps\del$ forces
$V_{cb}\sim\del$, while $V_{cb}\sim\del^2$
would have been more appropriate. Larger mixing
angles in model A.2 will result in values of
$m_c$ or $m_e$ that are too low. We show below
that the dangerous dimension-5 operators
inducing proton decay in models A and B are
suppressed by $\delta^w$. Hence, in models A.1,
A.2 and B.2 the leading LLLL and RRRR operators
are suppressed by ${m_e}/{m_d}$ relative to the
corresponding operators in minimal \suf and
thus fit, but marginally so, the constraints by
Ref.~\cite{Goto99,Murayama01}.

\subsection{Proton decay}
\label{proton_decay}
As we have seen above, triplet-matter couplings
and doublet-matter couplings originating from
higher dimension terms that involve some
bifundamentals, are distinguished by the
triplet symmetry. Consequently, the
triplet-matter Yukawa couplings can be
suppressed compared to minimal $SU(5)$. As we
will see, in models C, the triplet couplings
can also be enhanced compared to their $SU(5)$
values.

It is convenient to study the triplet couplings
by starting with 2-generation examples.
Consider then two generations with fields
$\tpr_1, \T_2, \fbarpr_1, \fbar_2$. The
triplet-matter couplings in models A and B are
then related to the down and up sector doublet
couplings, $y^u_{ij}$ and $y^d_{ij}$ by
\equas{
 \label{tripletDoubletSuppression_A_and B}
 \begin{array}{lr}
 y_{QQ} \sim
\left(
\begin{matrix}
y^u_{11}\delta^w & y^u_{12} \\
 y^u_{21}\delta^{w} & y^u_{22}\\
\end{matrix}
\right),
 &y_{e^c u^c} \sim
\left(
\begin{matrix}
y^u_{11}\delta^w & y^u_{12}\delta^w \\
 y^u_{21} & y^u_{22}\\
\end{matrix}
\right),\\ \\
 y_{u^c d^c}\sim \left(
\begin{matrix}
y^d_{11} & y^d_{12} \\
y^d_{21}\delta^w & y^d_{22}\delta^w\\
\end{matrix}
\right),&
y_{QL}\sim \left(
\begin{matrix}
y^d_{11} & y^d_{12}\delta^w \\
y^d_{21} & y^d_{22}\delta^w\\
\end{matrix}
\right).
\end{array}
}
Since model C has four massive triplets that
couple to matter, we have to consider not only
$y_{QQ},y_{e^c u^c},y_{Ql},y_{u^c d^c}$, the
couplings of $h_t$ and $\barh_t$ to matter, but
also $y_{QQ}^\prime,y_{e^c
u^c}^\prime,y_{Ql}^\prime,y_{u^c d^c}^\prime$,
the couplings of $\hpr_t$ and $\hbarpr_t$.
\equa{
 \label{tripletDoubletSuppression_C}
\begin{array}{ll}
y_{e^c u^c} \sim \left(
\begin{matrix}
y^u_{11}\delta^w & y^u_{12}\delta^w \\
 y^u_{21} & y^u_{22}\\
\end{matrix}
\right), &
y_{e^c u^c}^\prime \sim\delta^A \left(
\begin{matrix}
y^u_{11}\frac{\delta^w}{\eps} & y^u_{12}\delta^w \\
 y^u_{21} & y^u_{22}\eps\\
\end{matrix}
\right),\vspace{0.3cm}\\
y_{QQ} \sim \left(
\begin{matrix}
y^u_{11}\delta^w & y^u_{12} \\
 y^u_{21}\delta^{w} & y^u_{22}\\
\end{matrix}
\right),&
y_{QQ}^\prime \sim\delta^A \left(
\begin{matrix}
y^u_{11}\frac{\delta^w}{\eps} & y^u_{12} \\
 y^u_{21}\delta^{w} & y^u_{22}\eps\\
\end{matrix}
\right),
\vspace{0.3cm}
\\
y_{QL}\sim \left(
\begin{matrix}
y^d_{11}\delta^{-w} & y^d_{12} \\
y^d_{21}\delta^{-w} & y^d_{22}\\
\end{matrix}
\right),&
y_{QL}^\prime\sim\delta^A \left(
\begin{matrix}
y^d_{11}\frac{\delta^{-w}}{\eps} & y^d_{12}\frac{1}{\eps} \\
y^d_{21}\eps\delta^{-w} & y^d_{22}\eps\\
\end{matrix}
\right),\vspace{0.3cm}\\
y_{u^c d^c}\sim \left(
\begin{matrix}
y^d_{11} \delta^{-w}& y^d_{12}\delta^{-w} \\
y^d_{21} & y^d_{22}\\
\end{matrix}
\right),&
y_{u^c d^c}^\prime\sim \delta^A\left(
\begin{matrix}
y^d_{11} \frac{\delta^{-w}}{\eps}& y^d_{12}\frac{\delta^{-w}}{\eps} \\
\eps y^d_{21} & \eps y^d_{22}\\
\end{matrix}
\right).
\end{array}
}
Note that while in models A and B the dangerous
$QQQL$ and $e^c u^c u^c d^c$ dim-5 operators
are suppressed by $\delta^w$,  some of the
operators in model C seem to be a-priori
enhanced. The problematic couplings, say
$\left(y^\prime_{u^c d^c}\right)_{11}$, can
only be suppressed if $w$ is large enough so
that $\left(y^\prime_{u^c
d^c}\right)_{11}\sim\deln{n-k}$ with $k\ll n$.
This will serve as a strict requirement on
models C.2 and C.3 and will rule out model C.2.
\section{Results}
\label{results}
\label{examples}
Applying  our  results to specific models we
find that the models of the different classes
A, B, and C show qualitatively different
behavior. Models A.1 and A.2 are the only
models that necessarily break the unification
of lepton and down quark masses for a single
generation. The mass splitting, however, sets
$\del^w\sim 0.1$. The resulting flavor
parameters are viable in model A.1 but in model
A.2 this leads to $V_{cb}\sim 0.1$. The leading
dimension-5 operators, suppressed by
${m_e}/{m_d}\sim 0.1$, are consistent with
triplets at $\mgut$, but not lighter.

Models of class B (B.1, B.2, and B.3) are valid
for a wider range of $w$, allowing for a wider
range of flavor parameter and stronger
suppression of dimension-5 operators. For an
appropriate choice of $w$, these models also
allow triplets below than $\mgut$. In model
B.2, one can choose to partially break Yukawa
unification. With this choice, this model is
essentially the same as model A.1 in all other
respects.

In models C (C.2 and C.3) it is more difficult
to find a value of $w$ that satisfies both proton decay
and flavor constraints. In model C.2 this is impossible and
the model is ruled out (see Appendix). In model
C.3, however, such a choice is possible and
the model is viable. In this
model, too, the triplets may be made
lighter than $\mgut$.

In order to illustrate the above, we give one
example for each class.

\subsection{Model A.1}
This model has matter representations:
\equa{
\begin{array}{cccccc}
\tpr_{1}\ ,&\T_2\ ,&\T_3\ ,&\fbar_1\
,&\fbar_2\ ,&\fbar_3\ ,
\end{array}
}
and the Higgses are $h,\hbarpr$.

As discussed in section~\ref{e-d_splitting}, to
have $m_e/m_d\sim 0.1$, we must take $w=1$.
The requirement
$V_{us}\sim\sin{\theta_c}$, gives
$r=1$, and thus also $z=1$.
We then get
\equa{
m_u\sim \vev{h} \left( \begin{matrix}
 \epsilon \delta^3 & \epsilon^2 \delta^3 & \epsq \delta^2 \\
 \epsilon^2 \delta^2 & \delta^2 & \delta \\
\epsq \delta & \delta & 1
\end{matrix} \right),
\quad m_d \sim
\vev{\bar{h'}} \epsilon \left( \begin{matrix} \delta^4 & \delta^3 & \delta^2 \\
\delta^3 & \delta^2 & \delta \\
\delta^2 &\delta & 1
\end{matrix} \right),
}
and the charged lepton sector mass matrix
\equa{
\\m_l^T \sim
\vev{\bar{h'}} \epsilon \left( \begin{matrix} \delta^5 & \delta^4 & \delta^3 \\
\delta^3 & \delta^2 & \delta \\
\delta^2 &\delta & 1
\end{matrix} \right).
}
The resulting CKM matrix is
\equa{
V \sim  \left( \begin{matrix}
  1 & \delta & \delta^2 \\
  \delta  & 1 &  \delta \\
 \delta^2 & \delta & 1
\end{matrix} \right).
}

The dominant dimension-5 operators
$Q_1Q_1Q_2L_2$, $Q_2Q_1Q_1L_2$,
$u^c_1d^c_1u^c_2e^c_2,$ and
$u^c_2d^c_1u^c_1e^c_2$ are all suppressed by
$m_e/m_d\sim\delta$. Thus, the triplet mass can
be lower than the minimal \suf bound
\cite{Goto99} of $m_{h_t}\geq 7.2\times 10^{16}
\GeV$. However, we can not lower the triplet
mass to the range given in
Ref.~\cite{Murayama01} for coupling
unification.
\subsection{Model B.3}
\label{modelB3}
The matter field and Higgs
representations in this model are
\equa{
\begin{array}{cccccccc}
\tpr_1\ , &\tpr_2\ , &\T_3\ , &\fbarpr_1\ ,
&\fbarpr_2\ , &\fbar_3\ ,&h\ ,&\hbarpr\ ,
\end{array}
}

Let us set $ V_{cb} \sim \delta^2$ and
$m_c/m_t\sim \eps\delta$ so that $r=2,z=1$ and
hence $w=3$. We also take $V_{us}\sim\del$;
this determines  the up sector mass matrix. The
down sector mass matrix is determined by taking
$m_d\sim\deln{6},m_s\sim\deln{4}, m_b\sim\eps$.
The resulting mass matrices are
\equa{
\begin{array}{l}
\label{modelBmassMatrix}
m_u\sim \vev{h_d}\left(
\begin{matrix}
\epsilon \delta^3 & \epsilon \delta^2 & \epsq \delta^3 \\
 \epsilon\delta^2  & \eps \delta &\epsq \delta^2 \\
\epsq  & \epsq \delta^{n-1} & 1
\end{matrix} \right),\\ \\
m_d, m_l^T \sim \vev{\hbarpr_d} \left(
\begin{matrix}
\delta^6 & \delta^5 & \eps\delta^3 \\
\delta^5 & \delta^4 & \eps\delta^2 \\
\epsq\delta^3 &\epsq\delta^2  & \eps
\end{matrix} \right).
\end{array}
}
The CKM matrix is
\equa{
V \sim  \left( \begin{matrix}
  1 & \delta & \delta^3 \\
  \delta  & 1 &  \delta^2 \\
 \delta^3 & \delta^2 & 1
\end{matrix} \right).
}

The dominant dimension-5 operators,
$Q_1Q_1Q_2L_2$, $Q_2Q_1Q_1L_2$,
$u^c_1d^c_1u^c_2e^c_2,$ and
$u^c_2d^c_1u^c_1e^c_2$ are all suppressed by
$\deln{3}\sim 10^{-3}$. The strong suppression
of the dimension-5 operators in this model
allows for triplets as light as
$\order(\delta^2\mgut)$, compatible with gauge
coupling unification.
\subsection{Model C.3}
In this model the matter and Higgs fields are $\tpr_1, \tpr_2,
\T_3, \fbarpr_1, \fbarpr_2, \fbar_3, h, \barh$. In section
\ref{proton_decay} we have seen that to avoid enhanced dimension 5
operators we have to take $w$ as large as possible. We therefore
take $z=0$ (meaning that ${m_c}/{m_t}\sim\eps\delta^0$) and $r=2$
(corresponding to $V_{cb}\sim \delta^2$). The dangerous
triplet couplings are all suppressed if we take $A=1$ [see
equation \eqref{fourHiggsMassTerms}], and
\equas{
m_u\sim\vev{h_d}\matr{
\eps & \eps & \epsq\delta^2\\
\eps & \eps & \epsq\delta^2\\
\epsq\delta^{n-2}&
\epsq\delta^{n-2}&1},\vspace{0.4cm}\\
m_d,m_l^T\sim\vev{\barh_d}\matr{
\eps\delta^{2} & \eps\delta^{2} & \epsq\delta^{4}\\
\eps\delta^{2} &\eps\delta^{2} & \epsq\delta^{4}\\
\eps& \eps& \delta^{2}}\ .
}
For example, $\left(y_{QL}\right)_{11}\sim\deln{n-2}$ and
$\left(y_{QL}^\prime\right)_{11}\sim\deln{n-1}$. The first and
second generation terms may be split by introducing an additional
non-triplet, horizontal symmetry broken by a small parameter
$\eta$,
\equas{
\tilde{m}_u\sim\vev{h_d}\matr{
\eps\eta^2 & \eps\eta & \epsq\delta^2\eta\\
\eps\eta & \eps & \epsq\delta^2\\
\epsq\delta^{n-2}\eta&
\epsq\delta^{n-2}&1},\vspace{0.4cm}\\
\tilde m_d,\tilde m_l^T\sim\vev{\barh_d}\matr{
\eps\delta^{2}\eta & \eps\delta^{2}\eta & \epsq\delta^{4}\eta\\
\eps\delta^{2} &\eps\delta^{2} & \epsq\delta^{4}\\
\eps& \eps& \delta^{2}}.
}
Note that the quark masses and mixing angles
are all greater than or of the order of the
experimental parameters, so this model has
viable flavor parameters. However, the u-quark
mass, which is significantly larger than the
experimental value, $m_u\sim\eps\eta^2$ (note
that since $V_{us}\sim \eta$, then $\eta\gtrsim
0.1$), can not have its mass suppressed further
by a horizontal symmetry, triplet or
non-triplet, without suppressing $V_{us}$ below
its measured value.
\section{Naturally Light Triplets}
\label{lightTriplets} If the triplet
couplings to matter are small, we can try to
construct models with triplets below
$\mgut$, and still have an acceptable
proton-decay rate. In particular if the
triplets are around $10^{14}-10^{15}$ GeV
the standard model couplings unify
precisely~\cite{Murayama01} with no need for
any new thresholds or other corrections. To
construct such models, we can introduce a
new global symmetry $\tilde{Z}$, that
suppresses both the doublet and triplet
masses and leaves the fermion masses
unchanged. For example, this may be done for
models A and B, by taking
$\tilde{Z}(d^c)=-\tilde{Z}(\hbarpr)\neq 0$,
with all other fields neutral under
$\tilde{Z}$. In order to keep the doublets
at the electroweak scale, the
doublet-triplet mass ratio must be modified.
This can be easily accomplished by reducing
the size $n$ of the triplet symmetry.

However, we can also lower the triplet mass using the
triplet symmetry at hand. As we have already seen, models C can have triplets
at $\delta \cdot\mgut$.
To get lighter triplets in models A and B we must make sure
that the doublets remain light.

Consider for example the realization of model
B.3 of section~\ref{modelB3}. The Higgs masses
in the model came from $h\phibart\hbarpr+
h\phibard\hbarpr\deln{n-w}$, with $w=3$. Using
the triplet symmetry to suppress the triplet
mass by $\del^\alpha$ the superpotential Higgs
mass terms become
$h\phibart\hbarpr\deln{\alpha}+
h\phibard\hbarpr\deln{n-w+\alpha}$. Clearly, if
$\alpha\geq w$, the doublets become heavy.

It is interesting to note that in models A
and B the triplet symmetry can be used to
lower the triplet mass (while keeping the
doublets light) only in the models where the
dimension-5 operators are appropriately
suppressed. In these models, where the
leading dimension-5 operators are suppressed
by $ \delta^w$, the triplet symmetry can
suppress the triplet mass by up to
$\delta^{w-1}$; otherwise the doublets are
made heavy. This is the same condition that
we would have imposed thinking about proton
decay--- if the leading dim-5 operators are
to be suppressed by at least an order of
magnitude
${\delta^w}/{\delta^\alpha}\lesssim\delta$,
then $\alpha\leq w-1$.

Thus models with $w=1$, such as models A, can
not be viable with triplets below the GUT scale.
Models B.2 and B.3, however, can have triplets
at $\order(\delta^{w-1}\mgut)$, with dimension-5 operators
sufficiently suppressed and with light Higgs doublets.

\section{Conclusions}
\label{summary}
We studied the fermion masses and the Higgsino
mediated proton decay rate in \suff models. As
we explained, the two are tied together by the
triplet symmetry. We constructed a few examples
in which viable quark masses arise as a result
of the combination of the \suff gauge symmetry
and the triplet symmetry.

We also exhibited models in which, by virtue of the
large suppression of the proton decay rate, the
triplets can be below the GUT scale, and supply
the required threshold correction for gauge coupling unification.
As in all our models, the triplet mass in this case is
naturally obtained using the triplet symmetry.

\vskip 0.15in
{\noindent \bf Acknowledgements}
\vskip 0.125in
\noindent
We thank Yossi Nir for useful discussions.
Research supported by the Israel Science Foundation
(ISF) under grant 29/03,
and by the United States-Israel Science Foundation
(BSF) under grant 2002020.
\appendix

\section{model C.2}
\label{modelC2}
This model is ruled out by a combination of
constraints from flavor and proton decay. We
show below how this models parameters  have to
be chosen so that flavor is viable, and further
show that this choice leaves some triplet
couplings enhanced thus enhancing proton decay.

 In this model the generation
representations are
$\tpr_1,\T_2,\T_3,\fbarpr_1,\fbar_2,\fbar_3$.
The $\eps$-dependence of the mass matrices is
\equas{
m_u\sim\vev{h_d}\matr{
\eps & \epsq & \epsq\\
\epsq & 1 & 1\\
\epsq & 1 & 1
},
\qquad m_d\sim\vev{\barh_d}\matr{
\eps & \epsq & \epsq\\
\eps &1 & 1\\
\eps & 1& 1}.
}
Consider the $\tpr,\T$ mixing between the
first and second generations. The maximal value
of $r$ we can take is $r=1$, corresponding to
$V_{us}\sim\delta$. Note that both
$V_{us}\sim\delta$, and $V_{ub}\sim\delta^3$,
can only come from the mixing terms of the
first column in $m_d$, the rest of the mixings
are too small--- at least $\order(\epsq)$.
Thus, none of the terms of first column in $m_d$
should be of $\order(\deln{n-k})$ (for any
$k<<n$), and furthermore, the second and third
column terms should be at least of the same
order as the first column terms. Substituting
$r=1$, and taking the minimal order for the
second and third column terms, we are left with
three free parameters in the mass matrices
(which we denote by x,y,z),
\equas{
m_u\sim\vev{h_d}\matr{
\eps\delta^{2y+z} & \eps^2\delta^{2y+1} & \eps^2\delta^{y+1}\\
\epsq\delta^{2y+z-1} & \delta^{2y} & \delta^y\\
\epsq\delta^{y+z-1}&\delta^{y}&1
},
\vspace{0.4cm}\\
m_d\sim\vev{\barh_d}\matr{
\eps\delta^{x} & \epsq\delta^{x+2} & \epsq\delta^{x+2}\\
\eps\delta^{x-1} &\delta^{x+1} & \delta^{x+1}\\
\eps\delta^{x-1-y}& \delta^{x+1-y} & \delta^{x+1-y}}.
}
The Yukawa triplet couplings may be directly
related to the doublet Higgs couplings using
eq.~\eqref{tripletDoubletSuppression_C}. Since
$x,A\geq1$, we have to take $w=(2r-z)\geq 3$,
in order to suppress the $y_{QL}^\prime$ and
$y_{u^cd^c}^\prime$ first generation couplings.
This may only be done by taking $z<0$. We take
the minimal value of $z$ requiring
non-vanishing mass for the u-quark, $z=-2y$, so
$w$ is maximal. Requiring that
${m_d}_{31}\sim\eps\delta^0$, we get the
minimum value of $x$, $x=y+1$. Finally, in
order to avoid enhancement of
${y_{QL}^\prime}_{11}$ and
${y_{u^cd^c}^\prime}_{11}$ we set y=1. The
resulting mass matrices are
\equas{
m_u\sim\vev{h_d}\matr{
\eps\delta^{0} & \eps^2\delta^{3} & \eps^2\delta^{2}\\
\epsq\delta^{n-1} & \delta^{2} & \delta^1\\
\epsq\delta^{n-2}&\delta^1&1
},
\vspace{0.4cm}\\
m_d\sim\vev{\barh_d}\matr{
\eps\delta^{2} &\epsq\delta^{4} & \epsq\delta^{4}\\
\eps\delta^{1} &\delta^{3} & \delta^{3}\\
\eps\delta^{0} &\delta^{2} & \delta^{2}}.
}
Of the eight triplet Yukawa matrices, two still
have enhanced couplings,
\equas{
y_{u^c d^c}^\prime\sim \left(
\begin{matrix}
y^d_{11} \frac{\delta^{n-3}}{\eps}&
y^d_{12}\frac{\delta}{\eps\delta^4}
& y^d_{13}\frac{\delta}{\eps\delta^4}\\
y^d_{21} \eps\delta& y^d_{22}\eps\delta& y^d_{23}\eps\delta\\
y^d_{31}\eps\delta & y^d_{32}\eps\delta & y^d_{33}\eps\delta\\
\end{matrix}
\right), \vspace{0.4cm}\\

y_{QL}^\prime\sim \left(
\begin{matrix}
y^d_{11}\frac{\delta^{n-3}}{\eps} & y^d_{12}\frac{\delta}{\eps}
&y^d_{13}\frac{\delta}{\eps} \\
y^d_{21}\eps\delta^{n-3} & y^d_{22}\eps\delta &y^d_{23}\eps\delta\\
y^d_{31}\eps\delta^{n-3} & y^d_{32}\eps\delta & y^d_{33}\eps\delta
\end{matrix}
\right).
}
The decays through ${y_{QL}^\prime}_{12}$ and
${y_{u^cd^c}^\prime}_{12}$ are at least as
dangerous as the corresponding decays in
minimal \sufns. Thus, model C.2 can either have
viable flavor parameters or  have suppressed
dimension-5 operators--- it cannot have both!
We conclude that model C.2 cannot be made
viable.

\end{document}